\documentclass{article}
\usepackage{amssymb}

\begin{document}

\title{Statistical Structures Underlying Quantum Mechanics and Social
Science}
\author{Ron Wright\thanks{This article is based on a presentation at the Biennial Meeting of
the International Quantum Structures Association, Vienna, Austria,
July 2-8, 2002. I gratefully acknowledge helpful comments on a draft of this paper
from Charles Brainerd and David Foulis.}\\
University of Arizona, Department of Psychiatry\\ P. O. Box 245002, Tucson,
AZ 85724-5002, USA\\
\tt e-mail: wrightr@u.arizona.edu}
\date{}
\maketitle

\begin{abstract}
Common observations of the unpredictability of human behavior and the
influence of one question on the answer to another suggest social
science experiments are probabilistic and may be mutually
incompatible with one another, characteristics attributed to quantum
mechanics (as
distinguished from classical mechanics). This paper examines this
superficial similarity in depth using the Foulis-Randall Operational
Statistics
language. In contradistinction to physics, social science deals with
complex, open systems for which the set of possible experiments is
unknowable and outcome interference is a graded phenomenon resulting
from the ways the human brain processes information. It is concluded
that social science is, in some ways, ``less classical" than quantum
mechanics,
but that generalized ``quantum" structures may provide appropriate
descriptions of social science experiments. Specific challenges to
extending ``quantum" structures to social science are identified.
\end{abstract}

\section{Introduction}
Human beings are much bigger than electrons, and much more complex.
Humans do not change their states by small,
discrete amounts. And even though human beings themselves come in
discrete packages, their behavior typically does
not, at least at the level of analysis used in social science. All of
this suggests that descriptions of elementary particles
and descriptions of elementary school children are likely to be quite
different. I am not therefore suggesting that the same
phenomena occur in both, and yet, the roll of a die and proteins
crossing a cell membrane are also quite dissimilar, but
the language of ordinary statistics can successfully describe certain
aspects of both. In this article I examine ways in
which social science may require generalized statistical
understandings akin to understandings necessitated in physics
by the advent of quantum mechanics, as well as ways such social
science generalizations may be different from those
in quantum mechanics.

I offer the perspectives presented here as a practicing social
scientist: I have doctorates in both medicine and
psychology; I currently work in a department of psychiatry and have
spent a number of years in departments of
psychology, sociology and communication. I proffer these thoughts in
the context of my previous work in quantum logic
and Operational Statistics [1, 2, 3, 4, 5, 6]. One important goal is
to suggest possible directions for development, if social
sciences are to be encompassed by ``quantum" structures.

In order to highlight the parallels and contrasts between quantum
mechanics and social science I take specificÑ and
simplifiedÑversions of each. As representative of quantum mechanics I
take experiments on spin-one particles, wherein
states are unit vectors in the complex Hilbert space $\mathbb{C}^3$
and observables are Hermitian operators on that Hilbert space.
Obviously, this does not characterize all of ``quantum logic" or
``quantum structures"
and, indeed, I argue that ``quantum"
structures go beyond physics and provide, or at least suggest, the
requisites for understanding social science. This is
particularly true of the Foulis-Randall Operational Statistics
formulation, which I use in much of what follows.

For present purposes, I take social science to mean the study of
human behavior. As an illustrative example of social
science, I use studies of individual human beings in experimental
situations as might occur in cognitive psychology, with
basic experiments on memory providing specifics. I have tried to
describe below enough of the theory and
operationalization of the memory experiments to make the comparison
clear, but I have simplified both the theory and
the data somewhat in this presentation.

The outline of the paper is as follows: First I sketch reasons one
might expect similarities or differences between
quantum mechanics and social sciences. Then, to facilitate comparison
with social science, I describe the Hilbert space
manual of experiments on spin-one particles in the language of
Foulis-Randall Operational Statistics. I then review the
theory and operationalization of memory experiments in which subjects
learn lists of words from common categories
(e.g., fruits or animals), describing these also in the
Foulis-Randall language. Based on these expositions, I suggest ways
in which social science is similar to and different from quantum
mechanics at the level of mathematical and statistical
description. Finally, I outline what would be needed to extend
``quantum" structures to social science.

\subsection{Superficial similarities and differences}
A facile---and I think specious---argument for differences between
physics and social science is that the former is an
exact science while the latter is inexact. Physics is exact in that
it can predict the outcome of certain experiments with
great accuracy [but compare 7]. Consider, however, what happens if I
drop a feather. While the influences on the feather's
descent---gravity, rigidity and mass of the feather, air flow, air
temperature, initial linear and angular momentum,
distance to the floor, etc.---are all physical concepts, no one can, in
reality, predict exactly where the feather will fall.
Some might object that we could in principle do this if we made all
of the appropriate measures, but we cannot make all
these measurements for real feathers in real rooms. Yet we frequently
hold social science to the task of predicting how
a particular human being---a vastly more complex system than a feather
in a room---will behave in a certain situation.
This fundamental difference---the complexity of the phenomena to be
described and predicted---between physical and
social science is an underpinning of much of what follows here. The
task of learning words on a list is a very simple task,
yet it confronts us with many complications. Certain very low-level
tasks such as reaction times in attentional tasks might
be simpler (but maybe not) and most tasks of psychology are vastly
more complex than memory for word lists.

Nevertheless, there are also many reasons to believe that social
science might behave in a ``non-classical" way. A key
aspect of quantum mechanics is incompatible measurements. In social
science, asking one question of a human being
can change the state of the organism so that the answers to other
questions are altered. This could happen for external
reasons (e.g., so as to not appear inconsistent to the questioner) or
for internal reasons (e.g., one question might make
certain things more, or less, accessible in memory).

Incompatible measurements occur with spin-one particles when spin
components are measured along axes, say, 45
degrees apart, if the first measurement [8, p.72: ``measurements of
the first kind"] puts the system in an eigenstate of the
measurement operator. It may not be entirely dissimilar if a person
facing an emotional situation is first asked ``Do you
feel angry?" Answering ``Yes" may take the person from a blended
emotional state to an experience of anger, leading
to both another ``Yes" answer if asked again (say, by someone else)
and to changes in the responses to questions about
fear or sadness, a ``collapsing of the emotion wave," if you will.

And both situations may also be inherently probabilistic: Before the
first question, it may be neither true nor false that
the person is angry; they may be feeling a jumble of things that can
get conceptualized as anger or as something else.
This would seem, at least in some cases, to go beyond a ``mixed state"
consisting of 40\% anger, plus 35\% fear, plus....
There could also be an incompatibility in emotional states, so that
the more certain one is of being angry, the less certain
one is of being afraid, resulting in an ``uncertainty principle for
emotions."

With this sense that there are certain ``intriguing parallels" between
quantum mechanical phenomena and social science
phenomena, and yet important differences, let us begin an in-depth
comparison of specific examples, starting with a
common language. An ultimate question is whether Nature takes
advantage of the incompatibilities in social science
operations and produces results, as quantum mechanics does, that are
inconsistent with ordinary statistics.

\subsection{The Foulis-Randall Descriptive Language}
Foulis and Randall's Operational Statistics [e.g., 9, 10, 11] is a
generalization of the Kolmogorov [12, 13] description
of an experiment consisting of a set $X$ of possible outcomes and a $\sigma$-field
of subsets of $X$ identified as observable events;
probability measures on this $\sigma$-field are states of the system.

In the Foulis-Randall generalization, a set of experiments, or
``operations," is considered together and thought of as
the manual or handbook containing descriptions of all of the
operations that can be performed on the system under
consideration. Each operation is represented by its set of outcomes,
so a manual is a set of sets. These sets overlap if one
outcome can occur in two distinct experiments. Any subset of an
operation is considered an observable event. Events are
orthogonal if they operationally reject one another, that is, if they
are disjoint and their union is an event. A probability
measure, or weight function, $\omega$ for a manual is a mapping of outcomes
into $[0,1]$ with an unordered sum of 1 over any
operation; the set of all such weight functions is $\Omega$, the weight
space for the manual. An operational proposition for a
manual is defined via the set of outcomes that confirm the
proposition and the set of outcomes that refute the proposition.
Taken together, these ordered pairs of outcomes form the logic $\Pi$ for
the manual. Under favorable conditions $\Pi$ is an
orthomodular poset. The probability measures on the logic are called
states.

\section{A Quantum Mechanics Example: Spin-One Experiments}
Measurements on a spin-one particle correspond to Hermitian operators
on $\mathbb{C}^3$ and can be carried out using a
generalized Stern-Gerlach procedure [5]. The outcomes of the
corresponding operation are the eigenspaces of the
Hermitian operator, which are mutually orthogonal and span the space.
Hence, the manual of operations can be taken
to be all maximal orthogonal sets of subspaces of $\mathbb{C}^3$, or
equivalently, all maximal orthogonal sets of projections on $\mathbb{C}^3$
[{\it cf.} 14, Examples 12, 13]. Real physical measurements are presumed to
correspond approximately to these theoretical
measurements. Events are orthogonal sets of projections and two
events are orthogonal as events exactly when they are
orthogonal as sets of projections. A weight function on the manual is
a measure on the projection lattice and hence by
Gleason's Theorem [15] corresponds to a density operator on
$\mathbb{C}^3$.

The logic of the spin-one manual is isomorphic to the projection
lattice on $\mathbb{C}^3$: Outcomes that confirm the proposition
corresponding to projection $P$ are projections less than or equal to
$P$; outcomes that refute the proposition are projections
orthogonal to $P$. States on the logic are equivalent to weight
functions on the manual and therefore to density operators.
Note that the physical states go beyond the states on the logic in
that they contain phase information. States differing only
in their phase information correspond to the same weight function on
the manual, though they differ in the effect they
have in compound operations [3].

One potentially confusing aspect of the Hilbert space manual
described here is the multiple roles played by the same
mathematical object. Indeed, a projection onto a one dimensional
subspace plays four distinct roles: (1) it is an outcome,
(2) it is a proposition, (3) it is a state or weight, and (4) it is
an observable. The danger here is that, for example, a
superposition of states is mathematically equivalent to a
superposition of outcomes, propositions, etc.

\section{A Social Science Example: Memory Experiments}
\subsection{Memory Experiments as Social Science}
The study of human memory is a typical social science, and the study
of basic memory processes via experiments on
learning word lists would seem an uncomplicated example. Yet the
complexities are quite marked on both the theoretical
and applied sides.

Common sense suggests that events are remembered by storing a memory
trace in the brain and some memories are
stronger than others. The ``strength of the memory trace" should
predict memory performance. Research suggests that
this simple and intuitively appealing ``theory" is inadequate. In
particular, data on false memories---remembering things
that did not happen, say, remembering that you sent a letter when you
only ``wrote it in your head" or remembering false
lyrics to a song---suggest the need for a more complex theory of memory.

One such theory is Fuzzy Trace Theory, which asserts that people
remember things about the verbatim details of a
situation stored in a verbatim trace, but that they also abstract the
overall gist or meaning of the situation and separately,
and independently, store this gist or fuzzy trace; subsequent memory
performance is based on both traces [16, 17;
summarized in 18]. The verbatim trace has been found to fade more
quickly than the gist trace, so after a time, the exact
words spoken might be forgotten, but the meaning of what was said
remains: The
initial statement ``I was a math major in college" might be
remembered as
``I studied math in college" but not ``I learned hair styling in
beauty school." In the
type of memory experiment discussed below, subjects study a list of
words from specific categories. For example, they
might study HORSE, COW, DOG, APPLE, BANANA, PEAR. The verbatim trace
would be exactly this list of six
words. The abstracted gist might be ``animals and fruits." If a
subject remembers CAT or ORANGE, these are gist-consistent,
verbatim-false memories.

As in physics, there are controversies, subtleties and complexities
that this portrayal ignores, but it suffices to
exemplify a social science research paradigm. It is not important
whether Fuzzy Trace Theory is correct or even
plausible: It represents a typical social science theory, and its
ultimate fate depends on empirical support; assessing
whether the data support it requires an adequate statistical
description such as the one provided below.

\subsection{Recall, Recognition and the Fuzzy Trace Covert Judgments}
Memory tasks can be crudely divided into two types: Those that ask
you to recognize as a valid or invalid memory
something that is proposed to you (Think: Multiple-choice test), and
those that ask you to recall specific information
(Think: Fill-in-the-blank test). Both types of tasks can be and have
been used to study memory, and a viable memory
theory must explain both. In what follows I focus on recognition,
because it is conceptually and experimentally simpler.
Continuing the above word list example, a recognition test of memory
might ask
subjects to respond Yes (i.e., ``It was on the list.") or No (i.e.,
``It was not on the list.") to each of the following probe words:
HORSE, ORANGE, FLUTE,
CAT, HAMMER, PEAR. In this test list HORSE and PEAR are targets, that
is, words that were, in fact, on the list. The
remaining four words are distractors: CAT and ORANGE are related
distractors, while FLUTE and HAMMER are
unrelated distractors. (Note that actual study and test lists are
usually much longer than this---scores to hundreds of such
words from perhaps 6 to 20 categories.)

Fuzzy Trace Theory holds that, when a word is presented as a probe on
a recognition test, covert judgments are made
about the probe based on stored verbatim and gist traces, and these
covert judgments determine overt behavior, that is,
whether the probe is classified as a target, a related distractor or
an unrelated distractor. An {\it identity judgment} is made
if a sufficiently strong verbatim trace of a studied target matches
the probe. A {\it similarity judgment} is made if a sufficiently
strong gist trace of studied targets matches the probe. A {\it nonidentity
judgment} is made when, based on verbatim
information, a probe can be ruled out as a target: If the probe
ORANGE called into memory all three of the fruits on the
list, then ORANGE could be eliminated as a target. The likelihood of
these covert judgments differs for different probe
types. If both verbatim and gist traces are strong, identity and
similarity judgments would likely be made for targets, and
similarity and nonidentity judgments would likely be made for related
distractors; for unrelated distractors none of these
judgments is likely.

\subsection{The Theoretical Memory Manual}
``Systems are prepared" by having subjects listen to the words on the
study list. Characteristics of such systems can
be examined using various operations. Of interest for the moment are
operations asking subjects to identify probes as
targets ($T$), related distractors ($R$) or unrelated distractors ($U$).
Each operation has three possible outcomes and the
frequencies of these under various conditions are studied. The
prototypic operation is thus represented by the $TRU$
triangle shown in Figure 1. There is a different such
operation for each probe type. Hence the manual for the
experiments is $\{\{T_T, R_T, U_T\}, \{T_R, R_R, U_R\}, \{T_U, R_U,
U_U\}\}$ where $R_T$ is understood to be the outcome of the subject's
saying the probe was a related distractor when in fact it was a
target, etc.

\begin{center}
\begin{picture}(80,50)(0,0)
\put(20,0){\line(1,2){20}}
\put(40,40){\line(1,-2){20}}
\put(20,0){\line(1,0){40}}
\put(36.5,47){T}
\put(10,-10){R}
\put(64,-10){U}
\end{picture}

\vspace{.25in}
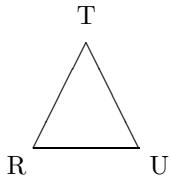
Figure 1: The $TRU$ Triangle
\end{center}

\bigskip
Already we are faced with a complication not faced by physics. If a
physical system is prepared and electrons are
bounced off the system to try to determine its properties, each of
the electrons
is identical. (Charley Randall: ``Electrons
don't have first names.") In the case of memory experiments, however,
various related distractors are not identical.
Indeed, asking about ORANGE vs. PERSIMMON will likely produce very
different responses, even though both are
fruits. Furthermore, not all targets are alike either: Although we
speak of the ``probability of an identity judgment for a
target probe," this probability differs from target to target,
depending, for example, on where the target occurred in the
study list and the memorability of the target itself.

To complicate matters further, the preparations are ``expensive" in
the sense that some human must sit and study the
words to be learned, then respond to the test. The supply of people
willing to do such things is often limited to the size
of a Psychology 101 class for a given semester. In light of this,
each ``system" (i.e., subject) is tested with multiple probes,
therefore, the actual experiment is a compound experiment. Past
research has shown that each probe disturbs the system,
so that the system is in a slightly different state after each probe.
Although these perturbations are occasionally
themselves of interest, more often they are nuisances that are
statistically averaged out, say, by randomizing the order
of the probes for different subjects, then collapsing responses
within probe type.

After collapsing, the manual for the experiment consists once again
of three
trichotomies $\{\{T_T, R_T, U_T\}$, $\{T_R, R_R, U_R\}$, $\{T_U, R_U,
U_U\} \}$, where $R_T$ is now understood to be an idealized outcome
whose frequency is the proportion of times the
subject said a target was a related distractor.

This idealized memory manual that is used in what follows, but it is
important to consider its arbitrary nature. If related
distractors came in two varieties, highly related (CAT, ORANGE) and
somewhat related (SKUNK, CORN), there would
be four probe types, hence four $TRU$ triangles, in the manual (or four
$TR_{hr}R_{sr}U$ rectangles, if subjects are to make this
discrimination). Since every possible probe is different from every
other, there are as many possible probe types as there
are words (and non-words are used in some variants), hence the
possible number of operations that could be carried out
on a preparation is large and unknown (and varies with time, plus
depends on the language spoken by the participant).
Moreover, the number of outcomes could be three or four or more, in
contrast to the spin-one manual where all (maximal)
operations have three outcomes (and to continua in physics where any
subdivision is possible, whereas for memory
research the ability to discriminate breaks down if levels of
relatedness are too close).

And worse yet, experiments are always carried out on ``mixed beams,"
since not
all ``preparations" are in the same
state: Some participants ``zoned out" as certain words were read;
some may not know some words (we try to avoid this
generally); some may not pay attention to the response instructions;
some may have emotional reactions to some fruits---i.e., like or dislike them---and hence
respond in an idiosyncratic way;
or some words may be more accessible because
they were used more recently. Changing one word on the study list, or
the order of the words, means the preparation is
different. We can do the same ``counterbalancing" act as for test
lists, but again there are very large number of
preparations and the relationship between two preparations is likely
to be ordinal ($B$ is between $A$ and $C$) rather than
metric ($A$ differs from $B$ by $x$ units on some scale). Furthermore,
different subjects abstract---and hence remember---different gists
(say, ``animals" vs. ``domestic animals" or ``fruits"
vs. ``edible plants").

\subsection{The Coarsened Memory Manual}
For technical and historical reasons, recognition experiments are
typically carried out as Yes-No experiments rather
than the three-way classification suggested in the last section.
Thus, the three-way classification $\{T, R, U\}$ is coarsened
into the three two-way classifications $\{T, T^{\prime}\}$, $\{R,
R^{\prime}\}$ and $\{U, U^{\prime}\}$, where $T^{\prime}$ means ``not
a target," etc. There are three
operations corresponding to each of these decision dichotomies, one
for each probe type, so the overall manual consists
of nine two-outcome experiments: $\{\{T_T, T_T^{\prime}\}$, $\{R_T,
R_T^{\prime}\}$, $\{U_T, U_T^{\prime}\}$, $\{T_R, T_R^{\prime}\}$, $\{R_R,
R_R^{\prime}\}$, $\{U_R, U_R^{\prime}\}$, $\{T_U, T_U^{\prime}\}$, $\{R_U,
R_U^{\prime}\}$,
$\{U_U, U_U^{\prime}\}\}$. As was true with the $TRU$ trichotomies above,
these operations are to be thought of as idealized operations
formed by compounding multiple probes in counterbalanced order and
collapsing by probe type. [It also turns out that
subjects often get confused if they are asked to change from one
discrimination decision to another, so in practice one
group of subjects makes all of the $T$ vs. $T^{\prime}$
classifications, another
group makes all the $R$ vs. $R^{\prime}$ decisions, and a third
group makes all the $U$ vs. $U^{\prime}$ decisions.]

A weight function on this manual is characterized by its value on the
first outcome of each dichotomy (column 1 of
Table 1). There are nine degrees of
freedom and $\Omega \equiv[0,1]^{9}$, since there is
no connection in the structure of the manual
between any two operations, hence no structural constraints on the
weight functions. (This is, of course, identical to the
situation for the $\mathbb{C}^2$ Hilbert space manual, and the reason
that Gleason's Theorem does not work there.) Similarly, the
logic of this semi-classical manual is nine copies of the Boolean
algebra 2 pasted together horizontally.

According to the simplified version of Fuzzy Trace Theory presented
here, the conditional empirical frequencies in
column 1 of Table 1 are determined by the probabilities of four
covert judgments:

\newpage
\parskip=0pt

\begin{itemize}
\item[$\iota_t=$] probability that an identity judgment is made for a
target probe
\item[$\sigma_t=$] probability that a similarity judgment is made for a
target probe given no identity judgment
\item[$\nu_r=$] probability that a nonidentity judgment is made for
a related distractor probe
\item[$\sigma_r=$] probability that a similarity judgment is made for a
related distractor probe given no nonidentity judgment
\end{itemize}

\noindent Column 2 of Table 1 gives the equations relating covert probabilities
to overt frequencies. The theoretically meaningful
states ({\it cf.} ``physical states") are ones that are generated by this
theoretical model. Via these equations, the nine-dimensional
weight space for the manual is reduced to a four-dimensional
theoretical subspace. The theory is evaluated
by collecting empirical weight function data (via compounding and
collapsing as above), then estimating the best
theoretical parameters to fit this data (iterative maximum likelihood
estimation), and finally comparing the empirically
observed frequencies with the frequencies predicted by the equations.
The predictions are generally within a couple
percentage points of the observed values [e.g., 19, Table 1].

\bigskip
\noindent Table 1: Probabilities assigned to different types of probes for each
of the three dichotomous discriminations.
$\omega_{x}(Y_{Z})$ is the probability that a $Z$ probe is called a $Y$ in the
$YY^{\prime}$ discrimination under conditions $x$.

\footnotesize
\begin{center}
\begin{tabular}{l|l|l|l|l}
\hline \parbox{.80in}{\begin{center} Theoretical Probabilty\end{center}}&
\parbox{.80in}{\begin{center} Probability in Terms
of the Theoretical\end{center}} & \parbox{.7in}{\begin{center} Perfect Gist \\ and Perfect\\
Verbatim\\
Memory ($\omega_{p}$)\end{center}} & \parbox{.7in}{\begin{center} No Gist
or\\ Verbatim\\ Memory
($\omega_{0}$)\end{center}} & \parbox{.7in}{\begin{center} Perfect Gist \\Memory and No
Verbatim\\ Memory ($\omega_{g}$)\end{center}}\\
\hline
$\omega(T_T) = P(``T"|T)$ & $\iota_t + \underline{(1- \iota_{t})\sigma_{t}}$ &
1 & 0 & 1\\
$\omega(R_T) = P(``R"|T)$ & $(1- \iota_{t})\sigma_{t}$ & 0 & 0 & 1\\
$\omega(U_T) = P(``U"|T)$ & $(1- \iota_{t})(1-\sigma_{t})$ & 0 & 1 & 0\\
\hfill Sum & \hfill$1 + \underline{(1- \iota_{t})\sigma_{t}}$ & \hfill 1 &
\hfill 1 & \hfill 2\\
$\omega(T_R) = P(``T"|R)$ & $\underline{(1- \nu_{r})\sigma_{r}}$ & 0 &
0 & 1 \\
$\omega(R_R) = P(``R"|R)$ & $\nu_{r} + (1- \nu_{r})\sigma_{r}$ & 1 & 0 & 1\\
$\omega(U_R) = P(``U"|R)$ & $(1- \nu_{r})(1- \sigma_{r})$ & 0 & 1 & 0\\
\hfill Sum & \hfill$1+ \underline{(1- \nu_{r})\sigma_{r}}$ & \hfill 1 & \hfill 1 &
\hfill 2\\
$\omega(T_U) = P(``T"|U)$ & 0 & 0 & 0 & 0\\
$\omega(R_U) = P(``R"|U)$ & 0 & 0 & 0 & 0\\
$\omega(U_U) = P(``U"|U)$ & 1 & 1 & 1 & 1\\
\hline
\end{tabular}

\parbox{4.75in}{Note: The Sum rows represents the total conditional probability for
the operation $\{T, R, U\}$ when these outcomes are
identified with the like-labeled outcomes in the dichotomies $\{T,
T^{\prime}\}, \{R, R^{\prime}\}$, and $\{U, U^{\prime}\}$. The
formulae in column 2 predict actual behavior; see text for definitions of the parameters.
The underlined terms represent the oddities of
behavior that cause the incompatibility of the operations; see text
for details. In this setup the four theoretical parameters
are estimated from the nine empirical probabilities. In real
experiments, subjects sometimes identify unrelated
distractors as targets or related distractors, for no obvious reason,
so three additional ``bias" parameters are needed, one
for each discrimination, resulting in seven parameters to be
estimated from nine data points [{\it cf.} 16].}
\end{center}

\normalsize

If memory is perfect, all four theoretical parameters are one, i.e.,
$(\iota_t, \sigma_t, \nu_r,$ $\sigma_r)$
$= (1, 1, 1, 1)$, and the weight function
$\omega_{p}$ is given by the third column in Table 1. If there is no memory
for the studied words, all four theoretical parameters
are zero, i.e., ($\iota_{t}, \sigma_t, \nu_r, \sigma_r)
= (0, 0, 0, 0)$, and
the weight function $\omega_0$ is given by the fourth column in Table 1. Since
memory for verbatim details fades more quickly than memory for the
overall gist, after a time delay starting from perfect
memory, $\iota_t$ and $\nu_r$ will be small, while $\sigma_t$ and
$\sigma_r$ will remain relatively large. The interesting phenomenon
occurs as one approaches the state of zero verbatim memory and perfect gist memory
as happens when the gist is strongly reinforced
(by, say, presenting many exemplars of the given category) and there
is a long delay between study and test, allowing
verbatim memory to fade. In this case $(\iota_{t}, \sigma_{t}, \nu_{r},
\sigma_{r}) = (0, 1, 0, 1)$, so, according to the above equations---and in agreement
with empirical data---the weight function is given by $\omega_{g}$ in the last
column in Table 1.

Behavior based on verbatim information is easy to predict and
understand; it is the way gist information is used that
produces the incompatibilities in the operations. Without verbatim
information, it is impossible to discriminate between
related distractors and targets, so how do people respond when a
similarity judgment is made, but neither an identity
judgment nor a nonidentity judgment is made? It depends on the
discrimination they are making. If asked whether a
probe is a target or not---the $T$ vs. $T^{\prime}$ discrimination---people accept
the probe as a target. If asked whether a probe is a
related distractor or not---the $R$ vs. $R^{\prime}$ discrimination---people accept
the probe as a related distractor. The term $(1-\iota_{t})\sigma_{t}$
therefore appears in both $P(``T"|T)$ and $P(``R"|T)$. Similarly, the
term $(1-\nu_{r})\sigma_{r}$ appears in both $P(``T"|R)$ and $P(``R"|R)$. If
the outcomes in the dichotomous discriminations are identified with
the corresponding outcomes in the three-way TRU
discrimination, then the conditional probabilities for each probe
type should add to one---but they do not, as indicated
in the Sum rows in Table 1. The sum exceeds one for targets by the
doubled term $(1-\iota_{t})\sigma_{t}$ and for related distractors by
the doubled term $(1-\nu_{r})\sigma_{r}$. If either verbatim information is perfect
(so $\nu_{t}= \nu_{r} = 1)$ or gist information is missing (so $\sigma_{t}=
\sigma_{r} = 0)$, then both terms are 0 and the sum over $TRU$ is 1, as
expected. In most real situations, the sum exceeds 1 and
equals 2 in the extreme situation represented by $\omega_{g}$.

Why might people would behave in this ``illogical" way? It is easy to
see how evolution might have favored this
behavior: Since much real-life memory occurs after a delay when
verbatim memory is weak yet gist is strong, we might
live our lives as ``something like that happened last week." Although
most people will back off to a position such as this
when pressed, normally we easily accept verbatim assertions as
correct if the overall Gestalt matches the remembered
gist sufficiently well. Think back to earlier in this paper: Did I
say ``I majored in math at college"?

The upshot of the behavior for the manual is (1) that the $TRU$
triangle cannot be added to the manual of dichotomies
(if the outcomes are identified) without losing empirically observed
weight functions, and (2) that coarsening the $TRU$
trichotomy to $\{T, T^{\prime}\}$ is not equivalent to coarsening $TRU$
to $\{T, \{R, U\}\}$: When the outcomes $R$ and $U$ are ``packed
together" to form $T^{\prime}$, they behave differently than the
event $\{R, U\}$, wherein the subject makes the $TRU$ discrimination,
but then $R$ and $U$ are lumped together for recording. A similar
phenomenon was noted by Tversky and Koehler [20,
abstract] in people's judgments of frequency: ``...judged probability
increases by unpacking the focal hypothesis and
decreases by unpacking the alternative hypothesis." The present
situation extends this finding beyond judgments of
frequency to acceptance of probes. Although packing $R$ and $U$ into
$T^{\prime}$ has none of the algebraic flavor of adding
projections in a Hilbert space, it is in some sense a ``coherent-like"
coarsening in that it behaves differently if we cannot
know which of the packed outcomes occurred, just as an interference
pattern is observed in the two-slit experiment if
we cannot know which hole the electron went through (ontologic
uncertainty), but not if we could know but did not
bother to look (epistemic uncertainty).

\section{Comparing Quantum Mechanics and Social Science}
\subsection{The Manual}
\begin{flushright}{\it In the empirical sciences, a well founded experimental program
ultimately is concerned with some cohesive
collection of physical operations---usually complete or exhaustive in
some sense.} Randall and Foulis [11, p.170]\end{flushright}

The manual of operations for spin-one particles is completely known
in the sense that the in-principle measurements
are represented by the self-adjoint operators on $\mathbb{C}^3$, and
the in-principle manual of elementary operations consists of all
maximal sets of mutually orthogonal $\mathbb{C}^3$ projections.
Indeed, every experiment already has a name, to wit, the set of
matrices representing the projections in the operation. Real life
measurements correspond to one of these in-principle
measurements in more or less known ways, and any in-principle
measurement can be realized by a suitable generalized
Stern-Gerlach (GSG) apparatus [5]. Although not all possible
real-life measurements are known, any such measurement
is presumed to be equivalent to one of these GSG measures via the
correspondence with Hermitian operators. Similarly,
the relationship between any two in-principle measurements, and hence
between any two real-life measurements, is
known, and the experiments are intricately intertwined. Any two
one-dimensional outcomes occur in overlapping
operations: The outcome orthogonal to both will be the overlap. (This
overlap is what constrains possible weight
functions and makes Gleason's Theorem work.) The manual of
in-principle operations---and up to GSG equivalence, the
manual of real-life operations---is exhaustive and coherent, as Randall
and Foulis required. The manual is parameterized
by the field of the GSG apparatus.

Experiments in the memory manual can also be changed in ``parametric"
ways paralleling rotations in space of the
Stern-Gerlach apparatus: (1) The preparation of systems can be
changed by using different numbers of exemplars of each
category; (2) The probes can be changed by varying their
prototypicality within the studied category; (3) the setting of
the discrimination can be changed by changing the delay between study
and test. Moreover, (4) the discriminations to
be made can be changed by including more than one level of
relatedness of distractors. There is therefore one memory
experiment for each preparation-probe-setting-discrimination
combination.

Although such combinations describe all experiments involving
studying words from categories and responding to
recognition probes, unlike the spin-one situation, what variations
are possible is not fully knowable. We know some
variations, but some differences once considered irrelevant are later
found to be important, introducing new variations.
If the experiments to be generalized to are ``memory for words from
categories," then the specific categories used should
not matter. Yet research has shown that concrete categories (e.g.,
fruits) and abstract categories (e.g., personality traits)
behave differently, so to handle all memory experiments, the
preparation needs an added dimension for abstractness of
the categories. Even beyond adding new dimensions there are problems
with identifying all possible preparations: What
constitutes a word? Is a word a particular participant has never
heard, a word for purposes of testing that participant's
memory? How long can a word list be? If it takes longer than a human
lifetime to listen to it, it is not a real experiment.
Furthermore, German word lists correspond to a different manual than
English word lists. In sum, the memory manual
is unknown and additions and refinements to the manual are
continually being made. In particular, the manual considered
by researchers at a given moment in time is not exhaustive.

There are theoretical relationships between measurements and outcomes
in memory research just as there are in
physics, but the experiments are not as intricately intertwined, and
relationships are generally ordinal rather than metric,
in contrast to the spin-one situations where rotating the GSG
apparatus by a certain amount has a precisely predicted
effect. For example, experimentally manipulating the number of
exemplars per category---say 5 vs. 3---will change the
memory traces, strengthening the memory trace for the category as a
whole (hence increasing $\sigma_t$ and $\sigma_r$) but having little
effect on $\iota_{t}$ for individual, once-presented targets. The effect
on $\sigma_t$ and $\sigma_r$ using 4 exemplars per category would be
between these two, but there is no prediction of exactly how much.
Hypotheses then typically have the form of ``such-and-
such manipulation causes an increase in a certain variable."

The relationship between real and idealized experiments in memory
research is also more complex than in physics.
While no physical measurement is perfect, the naive expectation is
that given a theoretical measurement, with sufficient
care, a real-life measurement as close as desired to the theoretical
measurement can be devised [but compare 21]. In
memory research, however, we wish to measure memory for categorized
lists in general, whereas all experiments use
specific categories that have properties not of theoretical interest.
Any real word-memory experiment is presumed to be
a poor representative of the theoretical word-memory experiment of
interest. In memory experimentation the implicit
theoretical manual is the set of experiments the present experiments
are presumed/claimed/intended to generalize to, but
the real manual uses a finite number of specific words.

We address this problem by using multiple specific categories and
averaging to eliminate the extraneous and retain
what results from common factors. A study of memory for fruits or
metals would be unconvincing as revealing general
memory phenomena, because the results might be influenced by
something specific about fruits or metals. It is only when
we demonstrate a phenomenon with multiple categories to show it is
independent of the categories chosen, that we have
convincing evidence about ``memory for categorized lists." The
reasoning here is similar to that of having multiple items
on an IQ test: Answering a particular item correctly is influenced by
IQ as well as other experiences and abilities; only
when we average many of these items together do the other things
``average out" leaving a more pure assessment of the
underlying IQ contribution.

\subsection{Theory and States}
In Operational Statistics, content-based theory is introduced in a
couple different ways. For both quantum physics and
social science, formal and informal theory goes into specifying the
manual, particularly outcomes that occur in the overlap
between operations. An important example is whether $T$ from $TRU$
should be identified with $T$ in $TT^{\prime}$. Theory introduced by
outcome identifications suffices for
spin-one experiments (but not spin-on-half), because
the structure of the manual determines exactly the weight functions
that correspond to physical states.

Since in memory research there is no {\it the} manual, its structure cannot
determine the states/weights (nor {\it the} logic), so
theory to specify the meaningful states must be added in some other
way. Fuzzy Trace Theory per above, specifies that
the meaningful states arise from specific values of $\iota_{t},
\sigma_{t}, \nu_{r}$,
and $\sigma_{r}$ via the equations in column 2 of Table 1.

States in social science are always partial states: Given the
complexity of a human being, we never suspect that we
have captured the full state, just as we would not believe we have
captured the full state of a feather in a room. In the
rarified realm of in-principle spin-one measurements, however, the
state of a specific particle (with phase information)
is a pure state, and is assumed to be the full and complete
description of the particle. For memory experiments, a full
description is not even desirable: Since we want to study memory for
categorized lists (say), we do not want a state that
specifies probabilities for each word, but rather for targets vs.
related distractors (say). In preparing spin-one particles,
each particle is in some pure state, with an ensemble of particles
forming a ``mixed beam" of known distribution. In
memory research, not only are all beams mixed (different subjects are
in different states), but each individual subject is
in a non-pure, partial state; we only have information about the
state with respect to certain specific experiences, allowing
multitudes of infinities of actual states, the specifics of which are
unknowable and not of interest.

\subsection{Coherent Coarsenings (Packing) and Outcome Interference}
\begin{flushright} {\it According to Feynman, the basic features of quantum mechanics are É a
probabilistic theory É outcomes that
interfere with each other É cannot be distinguished without
disturbing the system.} Gudder [22, p. xi]\end{flushright}

Memory experiments have the basic feature of quantum mechanics as
expressed in the above quote. Memory theories
are probabilistic theories. In the memory example above, the outcomes
$R$ and $U$ interfere with each other when they are
packed together as $T^{\prime}$ in the sense that the probability of
$T^{\prime}$ is not generally equal to the probability of the event
$\{R, U\}$,
which represents distinguishing between $R$ and $U$ experimentally.

The above quote continues with Feynman's belief that ``The main
feature of quantum mechanics is the way that
probabilities are computed" [22, p. xi], namely, taking the squared
modulus of complex amplitudes: If the coarsening
is coherent (outcomes interfere), the amplitudes are added and then
the squared modulus is taken; if the coarsening is
incoherent (no interference), the squared modulus is taken before
adding. Note that in computing these quantum
probabilities, there are exactly two versions: Add first or add
second.

If one takes the basic features of ``probabilistic theory with
interfering outcomes" to be the defining characteristic of
nonclassical theories (or general quantum structures), and
``computing probabilities via amplitudes" as defining a
specifically quantum theory, then memory theories are nonclassical
but not quantum. The packed coarsening of $R$ and
$U$ into $T^{\prime}$ resembles the ``coherent" coarsening in quantum
mechanics in that the probability is different from the
probability of the event $\{R, U\}$, but probabilities in this
situation cannot arise from amplitudes, even if one were to allow
amplitudes to come from some structure more general than $\mathbb{C}$ and have a
more general mapping into probabilities than
squared modulus. To see this one need only note a sense in which the
memory example is even less classical than
quantum mechanics: In comparing the experiments $\{T, T^{\prime}\}$
and $\{T,\{R, U\}\}$, not only are the probabilities of $T^{\prime}$ and
$\{R, U\}$ different, but (consequently) the probability of $T$ is
different in the two experiments, hence the probability of $T$
cannot arise from a underlying amplitude for $T$. If the spin-one
experiment $\{P1, P2, P3\}$ is coarsened into $\{P1, P2\oplus P3\}$
and $\{P1, \{P2, P3\}\}$, the probability of $P1$ remains unchanged.
Indeed, the probability of $P2\oplus P3$ is always the same as
the probability of $\{P2, P3\}$, because it is only in compound
experiments that the difference between coherent and
incoherent coarsenings shows itself [3].

The oddity in the memory experiments arises because of the
identification of
the outcome $T$ in $TRU$ with the $T$ in $TT^{\prime}$,
but that is how people seem to experience it. Yet empirically,
packing the outcomes $R$ and $U$ changes the identity of both
the packed $T^{\prime}$ and unpacked $T$ alternatives. So, subjects
experience $T$ as the same in $TT^{\prime}$ vs $TRU$, but assign it different
probabilities! The oddity, therefore, is in the way that human brains
process information, rather than (at least in any direct
way) a macroscopic manifestation of some true quantum phenomenon [{\it cf.}
23]. This also puts the locus of the oddity in
a vastly more complex system (the brain) that is likely to behave in
more unruly ways than spin-one particles.

That humans are making judgments also has ramifications for compound
experiments. It might be of interest, for
example, to know what would happen if the $TT^{\prime}$ experiment
was followed by the $TRU$ experiment. If the classification
is to be made about the same probe, the question becomes ``You said
HAMMER was a target, now say whether it is a
target, a related distractor or an unrelated distractor." The
subject's response is likely to be: ``I just told you it is a target."
But moreover, once the $TRU$ discrimination is introduced, it is
likely to unpack $R$ and $U$ in the subject's mind, changing
subsequent judgments in the $TT^{\prime}$ discrimination. This is a
further sense in which the $TT^{\prime}$ and $TRU$ experiments are
incompatible. And $TT^{\prime}$ and $RR^{\prime}$ may be incompatible
for similar reasons: They imply the ``common refinement" $TRU$,
thereby unpacking $T^{\prime}$ and $R^{\prime}$ and changing the
original dichotomies. It is exactly this sense in which $TRU$ is not a
common refinement of $TT^{\prime}$ and $RR^{\prime}$. It is also
interesting to note here that unpacking might not be complete,
allowing for intermediates between the fully packed outcome $T^{\prime}$ and
the fully unpacked event $\{R, U\}$, in contrast to the
dichotomous nature of coarsenings in quantum mechanics, i.e., whether
you take the modulus before or after adding the
amplitudes.

Given the influences on human thinking, it may not be possible to do
any compound operations on the same probe,
in stark contrast to physics where a particle coming out of one
Stern-Gerlach apparatus can be fed unproblematically into
a second. Indeed, the physical theory explaining one such experiment
generally allows us to predict the outcomes of compound experiments without
having a specific theory of them. This
generalization in physics from simple to
compound operations is often based on the idea that measurements are
of the ``first kind" [8]. Are memory measurements
of the first kind? The ``I just told you...." response suggests they
might be, but the resulting state is not pure as it often
is in physics, and if, as just discussed, compound operations in
memory research is problematic, how can we base the
definition on empirical data?

\subsection{Other Issues}
There are several other issues worth considering briefly because they
are often raised as prototypic of quantum
mechanics or social science. For example: Do superpositions of states
occur in social science? If a superposition is a
linear combinations of vectors in a complex state space, then this
question is meaningless, since there are no complex
state spaces for social science experiments. If the question is about
some generalization of the concept of superposition
such as the one proposed by Foulis, Piron and Randall [14, p.823],
then certainly these exist in social science; indeed
$\omega_{p}$ is a superposition in this sense of $\omega_{0}$ and
$\omega_{g}$, as is any
weight that treats unrelated distractors as such, no matter what
it does for targets and related distractors. What is less clear is
whether this tells us anything useful.

Another difference between quantum mechanics and social science,
related to complexity, is the question of isolated
systems. In both quantum and classical mechanics, we perform Gedanken
experiments on isolated systems and fancy
that we can create a system as nearly isolated as needed to check out
the theory. Social science deals with open, non-isolated
systems. In the memory experiments, for example, subjects come to the
experimental sessions with a lifetime
of using the words that they will be tested on, and recent usage
outside the lab may parallel studying the word, so the
outcome of the experiment is influenced by extra-experimental factors
which can never be eliminated. As was the case
with having to use specific words to test types of words, we dampen
extra-experimental influences by averaging over
many subjects, no one of which is an isolated system.

Both physics and social science use mathematical models---e.g., the
Hilbert space model or the Fuzzy Trace model---but the understanding of
what the models represent is different. In
physics, a model is taken to represent reality which
actual experiments approximate, and although there are different
Hilbert space models for different systems, a given
system has a specific mathematical model. A physical state is a full
description of the system at a given moment. In social
science, models are generally not representative of reality, but
merely descriptive: Things work out as if this were true.
If a model is empirically supported, any theory must be consistent
with this model to be credible. Fuzzy Trace Theory
does not specify a specific mathematical model for memory
experiments, but rather offers principles for how to construct
models for specific situations. The model presented above provides
good predictions for the memory of words in
categories, but if the situation is changed---e.g., if there were two
levels of relatedness of distractors---then Fuzzy Trace
Theory would add other parameters.

\section{Summary and Agenda for Future Research}
It is all too common these days for people to add ``quantum" to a
concept just for the cachet of the word, for example
``quantum healing" (healing that occurs in tiny quanta?). That is not
my intention here. I am not suggesting that social
science is quantum like. Rather, I am suggesting that quantum
mechanics has forced us to consider nonclassical logic
and statistics, and that social science is even more nonclassical
than quantum mechanics.

Quantum mechanics and social science are similar in that they can
both be described by the Foulis-Randall Operational
Statistics model, they are both inherently probabilistic (for
different reasons), they both deal with incompatible
experiments and they both exhibit outcome interference. And we have
seen that Nature does ``take advantage of" the
incompatible experiments by frequencies incompatible with a common
refinement.

Yet, quantum mechanics and social science are also fundamentally
different in that quantum mechanics provides a rich, complete mathematical
description of simple, isolated systems
resulting in a small set of physical laws about a manageable number of elementary
particles and forces, whereas social
science deals with sparse, partial descriptions of complex, open systems, resulting
in many approximate mini-theories
about vast numbers of ``objects," for example, the unknown and unknowable number
of words in the memory experiments.

Fuzzy Trace Theory in particular has much in common with quantum
theory. Both began with facts that did not make
sense given understandings at the time (e.g., interference patterns
in the two-slit experiment or acceptance of related
distractors as targets in memory research). Both proposed
unobservable mathematical features (amplitudes in physics,
probabilities of covert judgments in memory) and equations that
predict observed phenomena from these unobservable
features. In both domains, theories sink or swim based on whether
they predict observations across situations. Both
predict that states evolve over time, based on the unobserved
features, and each in its characteristic way: Quantum
mechanics specifies a Hamiltonian describing metrically how the state
evolves; Fuzzy Trace Theory specifies ordinally
how parameters change with delay between study and test.

The similarities between quantum mechanical and social scientific
theories make it tempting to take mathematical
structures derived to describe quantum structures and apply them, or
adapt them, to the social sciences. This paper
suggests, however, specific challenges to researchers who desire to
build a foundation---as Randall and Foulis (for
example, [11], as quoted in section 4.1) proposed to do---for all empirical
science. I list some key challenges here:

\begin{enumerate}
\item We cannot assume that the manual of all relevant experiments is
known, and if the manual is not exhaustive, we
cannot judge its coherence or other similar properties. Without
knowing the manual, we cannot know the logic,
so cannot ask about its properties, for example, is the logic an
orthomodular poset?
\item Theoretical considerations have to be brought in exogenously to
the structure of the manual itself, for example,
via the parameters and equations derived from Fuzzy Trace Theory.
\item Outcome interference is less neat in social science than in
quantum mechanics (for example, the graded packing
of $R$ and $U$ described above), occurs for different reasons (namely the
way human brains work rather than the
nature of the universe directly), and may change frequencies of
non-coarsened outcomes.
\item The relation of real experiments to theoretical experiments is
more complex in social science, for example,
having to test memory for categorized words using specific words from
specific categories. To approximate
theoretical outcomes we must average over multiple subjects, study
lists and probes, none of which is in itself
a particularly good example of the entity of interest. (One may
average in physics, too, but often over replications
of the same experiment.)
\item Theoretical relationships and empirical hypotheses are mostly
ordinal, not metric. For example, using a longer
study list might be predicted to change overall memory
performance, but not by a specific amount.
\item Mathematical models in social science are approximations to
real data, not truth. States are partial states on open
systems.
\end{enumerate}

The dismissive view of psychology as soft science leads to the
spurious question: Can psychology become like physics
when it grows up? I hope that the above thoughts can allow physicists
to move beyond this stereotypic view of social
sciences as merely sloppy physics, and can offer those working in
quantum structures a vision of the territory that must
be traversed before a true description of empirical science is in
hand. And that leaves the final question of whether the
new description can do real work in social science.


\begin{thebibliography}{23}
\bibitem{} Wright, Ron (1977). Projection-valued states. Ph.D. dissertation at
the University of Massachusetts/Amherst.
\bibitem{} Wright, Ron (1977). The structure of projection-valued states: A
generalization of Wigner's theorem. International
Journal of Theoretical Physics, 16, 567-573.
\bibitem{} Wright, Ron (1978). Spin manuals: Empirical Logic talks quantum
mechanics. In A. R. Marlow (Ed.), Mathematical
foundations of quantum theory (pp. 177-254). New York: Academic Press.
\bibitem{} Wright, Ron (1978). The state of the pentagon: A nonclassical
example. In A. R. Marlow (Ed.), Mathematical
foundations of quantum theory (pp. 255-274). New York: Academic Press.
\bibitem{} Swift, Arthur R., \& Wright, Ron (1980). Generalized Stern-Gerlach
experiments and the observability of arbitrary
spin operators. Journal of Mathematical Physics, 21, 77-82.
\bibitem{} Wright, Ron (1990). Generalized urn models. Foundations of Physics,
20, 881-903.
\bibitem{} Hedges, Larry V. (1987). How Hard Is Hard Science, How Soft Is Soft
Science? The Empirical Cumulativeness of
Research. American Psychologist, 42, 443-455.
\bibitem{} Pauli, Wolfgang (1958). Die allgemeinen Prinzipien der
Wellenmechanik. Handbuch der Physik, Band V, Teil 1,
1-168.
\bibitem{} Foulis, David J., \& Randall, Charles H. (1972). Operational
Statistics I: Basic concepts. Journal of Mathematical
Physics, 13, 1667-1675.
\bibitem{} Randall, Charles H., \& Foulis, David J. (1973). Operational
Statistics II: Manuals of operations and their logics.
Journal of Mathematical Physics, 14, 1472-1480.
\bibitem{} Randall, Charles H., \& Foulis, David J. (1978). The operational
approach to quantum mechanics. In C. A. Hooker
(Ed.), Physical theory as logico-operational structure (pp. 167-201).
Boston: D. Reidel.
\bibitem{} Kolmogorov, A.N. (1933). Grundbegriffe der
Wahrscheinlichkeitsrechnung. Ergebnisse der Mathematik, 2, Berlin.
\bibitem{} Kolmogorov, A.N. (1956). Foundations of the theory of probability
(2nd ed). New York: Chelsea.
\bibitem{} Foulis, David J., Piron, Constantin, \& Randall, Charles H. (1983).
Realism, operationalism and quantum mechanics.
Foundations of Physics, 13, 813-841.
\bibitem{} Gleason, Andrew M. (1957). Measures on the closed subspaces of a
Hilbert space. Journal of Mathematics and
Mechanics, 6, 885-893.
\bibitem{} Brainerd, C. J., Reyna, V. F., \& Mojardin, A. H. (1999). Conjoint
recognition. Psychological Review, 106, 160-179.
\bibitem{} Brainerd, C. J., \& Reyna, V. F. (2001). Fuzzy-trace theory:
Dual-processes in reasoning, memory, and cognitive
neuroscience. Advances in Child Development and Behavior, 28, 49-100.
\bibitem{} Brainerd, C. J., \& Reyna, V. F. (2002). Fuzzy-trace theory and
false memory. Current Directions in Psychological
Science, 11, 164-169.
\bibitem{} Brainerd, C. J., Wright, Ron, Reyna, V. F., \& Mojardin, A. H.
(2001). Conjoint recognition and phantom
recollection. Journal of Experimental Psychology: Learning, Memory, \&
Cognition, 27, 307-327.
\bibitem{} Tversky, Amos, \& Koehler, Derek J. (1994). Support theory: A
nonextensional representation of subjective
probability. Psychological Review, 101, 547-567.
\bibitem{} Foulis, David J., \& Gudder, Stanley P. (2001). Observables,
calibration, and effect algebras. Foundations of Physics,
31, 1515-1544.
\bibitem{} Gudder, Stanley P. (1988). Quantum probability. San Diego:
Academic Press.
\bibitem{} Hagan, S., Hameroff, S. R., \& Tuszynski, J. A. (2002). Quantum
computation in brain microtubules: Decoherence
and biological feasibility. Physical Review E. Statistical,
Nonlinear, \& Soft Matter Physics, 65.
\end{thebibliography}
\end{document}